# Thermonuclear Ignition of Dark Galaxies

J. Marvin Herndon
Transdyne Corporation
San Diego, CA 92131 USA

April 13, 2006

## Abstract

Dark matter is thought to be at least an order of magnitude more abundant than luminous matter in the Universe, but there has yet to be an unambiguous identification of a wholly dark, galactic-scale structure. There is, however, increasing evidence that VIRGOHI 21 may be a dark galaxy. If VIRGOHI 21 turns out to be composed of dark stars, having approximately the mass of stars found in luminous galaxies, it will pose an enigma within the framework of current astrophysical models, but will provide strong support for my concept, published in 1994 in the Proceedings of the Royal Society of London, of the thermonuclear ignition of stars by nuclear fission, and the corollary, non-ignition of stars. The possibility of galactic thermonuclear ignition is discussed from that framework and leads to my suggestion that the distribution of luminous stars in a galaxy may simply be a reflection of the galactic distribution of the heavy elements.

Although dark matter is thought to be at least an order of magnitude more abundant than luminous matter in the Universe, there has yet to be an unambiguous identification of a wholly dark, galactic-scale structure. There is, however, increasing evidence that VIRGOHI 21, a mysterious hydrogen cloud in the Virgo Cluster, discovered by Davies et al. [1] may be a dark galaxy. Minchin et al. [2] suggested that possibility on the basis of its broad line width unaccompanied by any responsible visible massive object. Subsequently, Minchin et al. [3] find an indubitable interaction with NGC 4254 which they take as additional evidence of the massive nature of VIRGOHI 21. If their identification of VIRGOHI 21 as a dark galaxy is validated, that discovery will open a new era in astrophysics.

If VIRGOHI 21 turns out to be composed of dark stars having approximately the mass of stars found in luminous galaxies, it will pose an enigma within the framework of current astrophysical models. Popular stellar formation models, for all but brown dwarfs, are based upon assumptions leading to the certainty of luminosity. But popularity only measures popularity, not scientific correctness. In a paper published in 1994 in the *Proceedings of the Royal Society of London*, I suggested reasons for the possible existence of solar-mass-size dark stars [4]. From that theoretical standpoint, it seems appropriate to speculate on the implications that would arise if indeed VIRGOHI 21 is proven to be composed of non-brown-dwarf, full-size dark stars.

At the beginning of the 20$^{th}$ Century, one of the most pressing astrophysical questions concerned the nature of the energy source that powers the Sun and other stars. For a time it was thought that the heat arose from the in-fall of dust and gas during formation. But calculations ultimately showed that such heat was insufficient to power the Sun for as long as life has existed on Earth. Then, thermonuclear fusion was discovered [5]. By 1938, thermonuclear fusion reactions, thought to power the Sun and the other stars, were reasonably well understood [6, 7]. The million degree temperatures necessary to initiate thermonuclear fusion reactions were assumed to arise from gravitational collapse of dust and gas during star formation [8, 9].

In the 1960s, when computers first became readily available, Hayashi and Nakano [10] discovered that the collapse of dust and gas was insufficient to attain thermonuclear ignition temperatures of about a million degrees. The reason is that surface heating from the in-fall of dust and gas is offset by radiation from the surface, which is a function of the fourth power of temperature. Subsequent stellar formation models thus invoked the assumption of an additional shock-wave-induced sudden flare up or tweaked parameters such as accretion rate and opacity in attempts to attain thermonuclear-ignition temperatures [11]. One might expect from observations of nature that a common, widespread event would admit a broad domain of formative conditions, rather than highly-specific, special circumstances.

In 1969, astronomers discovered that Jupiter radiates to space more energy than it receives. Verification followed, indicating that not only Jupiter, but Saturn and Neptune as well each radiate approximately twice as much energy as they receive from the Sun [12, 13]. For two decades, planetary scientists could find no viable explanation for the internal energy sources in these planets and declared that "by default" [14] or "by elimination" [15] the observed energy must come from planetary formation about 4.5 x 10$^9$ years ago.

In 1992, using Fermi's nuclear reactor theory, I demonstrated the feasibility for planetocentric nuclear fission reactors as the internal energy sources for the giant outer planets [16]. Initially, I considered only hydrogen-moderated thermal neutron reactors, but soon demonstrated the feasibility for fast neutron reactors as well, which admitted the possibility of planetocentric nuclear reactors in non-hydrogenous planets [4, 17, 18]. Subsequently, these calculations were verified [19] and extended with state-of-the-art numerical simulation calculations [20, 21] yielding results consistent with geophysical observables [22, 23].

The concept of planetocentric nuclear fission reactors may also apply to protostars and forms the basis for my suggestion that thermonuclear fusion reactions in stars, as in hydrogen bombs, are ignited by self-sustaining, neutron-induced, nuclear fission [4]. The corollary to thermonuclear ignition is non-ignition, which might result from the $^{235}$U/$^{238}$U ratio being too low, or from the absence of fissionable elements, and which would lead to dark stars.

Observational evidence, primarily based on velocity dispersions and rotation curves, suggests that spiral galaxies have associated with them massive, spheroidal, dark matter components, thought to reside in their galactic halos [24]. Interestingly, the luminous disc stars of spiral galaxies belong to the heavy-element-rich Population I; the luminous spheroidal stars of spiral galaxies belong to the heavy-element-poor Population Il. In spiral galaxies, the dark matter components are thought to be associated in some manner with the spheroidal heavy-element-poor Population II stars [25, 26]. The association of dark matter with heavy-element-poor Population II stars is inferred to exist elsewhere, for example, surrounding elliptical galaxies [27, 28]. Because of the apparent association of dark matter with heavy-metal-poor Population II stars, I have suggested the possibility that these dark matter components are composed of what might be called Population III stars, zero metallicity stars or stars at least devoid of fissionable elements, and, consequently, unable to sustain the nuclear fission chain reactions necessary for the ignition of thermonuclear fusion reactions.

The discovery of full-size dark stars would lend strong support to my concept of stellar thermonuclear ignition by nuclear fission. A dark galaxy composed of such dark stars would have profound implications. Although VIRGOHI 21 has not yet been shown to be a dark galaxy, the evidence to date looks promising and so warrants forethought as to potential implications.

The existence of a dark galaxy composed of non-brown-dwarf, full-size dark stars would certainly call into question the long-standing idea of gravitational collapse as the sole source of heat for inevitable stellar thermonuclear ignition, which after all has no laboratory support, unlike my idea [4] of a nuclear fission trigger. Population III stars comprising a dark galaxy might well be considered first-generation stars, devoid of metallicity. But, such a concept, although understandable, may lead to a seeming enigma, a paradox within the framework of current astrophysical models.

For half a century, the concept that elements are synthesized within stars [29] has become widely accepted. In the $B^2FH$ model, heavy elements are thought to be formed by the R-process at the end of a star's lifetime. If actinide elements are required to ignite stars, but are formed at the end of a star's lifetime, then how did the first stars ignite? The solution to that paradox may be that something like an R-process occurs in other circumstances. I suggest that heavy elements are formed by gravity-induced, massive explosions of nuclear matter at the galactic core. These heavy elements are ejected into the surrounding galaxy, where they tend to settle in the galactic plane, and seed dark stars, subsequently igniting stellar thermonuclear fusion reactions, and changing dark galaxies into luminous ones. From that perspective, the distribution of luminous stars in a galaxy may simply be a reflection of the distribution of the heavy elements.

Dark galaxies comprised of dark stars have not yet been discovered. But when they are, it will certainly be one of the most important astrophysical discoveries of the century.

Within the context of the ideas presented here, that discovery will pose no enigma and no paradox.